\documentclass[aps,prl,twocolumn,amsmath,amssym,amsthm,superscriptaddress,reprint]{revtex4-1}
\pdfoutput=1
\usepackage[colorlinks=true,
linkcolor=blue,
urlcolor=blue,
citecolor=blue]{hyperref}
\usepackage{graphicx}
\usepackage{dcolumn}
\usepackage{bm}
\usepackage{url}
\usepackage{bm}   
\usepackage{bbm}   
\usepackage{verbatim}
\usepackage{stmaryrd}
\usepackage{amsthm}
\usepackage{amssymb}
\usepackage{array,multirow}
\usepackage{empheq}
\usepackage[usenames,dvipsnames]{color}
\usepackage[normalem]{ulem}
\usepackage{tikz}

\newcommand{\LS}{Szil\'{a}rd}

\DeclareMathOperator{\sign}{sign}

\begin{document}
\title{Thermodynamically rational decision making under uncertainty}
\author{Dorian Daimer}
\affiliation{Department of Physics and Astronomy \\ University 
of Hawaii at M\=anoa, Honolulu, HI 96822, USA}

\author{Susanne Still}
\email[Corresponding author: ]{sstill@hawaii.edu}
\affiliation{Department of Physics and Astronomy \\ University 
of Hawaii at M\=anoa, Honolulu, HI 96822, USA}

\begin{abstract}
An analytical characterization of thermodynamically rational agent behaviour is obtained for a simple, yet non--trivial example of a ``Maxwell's demon" operating with partial information. Our results provide the first fully transparent physical understanding of a decision problem under uncertainty.
\end{abstract}
\maketitle

Decision making under uncertainty is one of the core hallmarks of intelligence, for which conclusive theory is yet outstanding. 
Mathematical foundations are commonly phrased in terms of optimization. 
Notions of ``utility" \cite{neumann1947theory, horvitz1988decision}, ``reward" \cite{sutton2018reinforcement}, ``risk" \cite{kahneman1984choices} and ``regret" \cite{loomes1982regret} are, for example, used to formalize agent objectives and define rationality. 
But the exact quantification of these subjective notions typically expresses {\it ad hoc} modeling choices, ultimately based on the designer's intuition.
This approach can be problematic, particularly when complex behaviors are modeled and model performance measures are not unambiguous.

An alternative is to look for {\em physical} foundations. All decision makers are physical systems, and physical limits to information processing can provide design rules \cite{CB}. 
As with many complex problems, a first step towards building foundational understanding is to carefully study a rudimentary base case. The simplest decision problem is a binary choice. 

Interestingly, Maxwell's demon \cite{maxwell-demon}, embodies a binary decision problem in which no {\it ad hoc} assumptions are necessary to quantify rationality. The observer's goal is unambiguous: available information is used to harvest energy. The ``demon" can be understood as an agent operating an information engine \cite{szilard-german}.
The energy required to take decisions must be subtracted from the derivable gain, and the bottom line is the engine's remaining work output. A thermodynamically rational agent chooses that strategy which allows for the largest net output \cite{CB}. 

In recent years, numerous experiments have verified 
that microscopic information engines are feasible 
\cite{toyabe2010experimental, berut2012experimental, koski2014SEszilard, jun2014high, vidrighin2016photonic, hong2016experimental, gavrilov2016erasure, gavrilov2017direct, chida2017power, cottet2017observing, masuyama2018information, ribezzi2019large, paneru2020colloidal, paneru2020efficiency, dago2021information, koski2014experimental, koski2015chip, camati2016experimental, paneru2018lossless, admon2018experimental, saha2021maximizing} and potentially useful in real world implementations \cite{saha2023information}, such as biomolecular machinery in cells \cite{flatt2023abc}. This experimental evidence has lifted Maxwell's demon and \LS's information engine from the realm of thought  experiments, useful for understanding the physical nature of information, to the realm of potential applications in energy efficient computing.

Real world agents and systems make decisions based on observable data, which may contain only {\em partial} information about the variable (or multiple variables) required to decide which actions to take. These relevant variables have to be {\em inferred} from observations. Typical real world decision problems thus contain intrinsic uncertainty due to partial observability. The framework of generalized, partially observable information engines affords a principled derivation of the observer's inference method from physical arguments \cite{CB}. 
But even simple examples of partially observable \LS\ engines have nontrivial solutions that are not analytically tractable \cite{stilldaimer2022}.

Here we present an example that can be understood analytically. Motivated by the fact that quite commonly not all outcomes of an observable carry useful information for the purpose of making a decision, we consider the simplified case in which observations either confer full, uncorrupted information about the variable of interest, or they are entirely uninformative about it. 
We show that this type of problem can be mapped onto a partially observable \LS\ engine. 
Thermodynamically rational decision policies are presented in a parameterized form which affords clear interpretability and analytic analysis. They differ from intuitive coarse graining of the observable, a commonly used method \cite{gibbs1902elementary}. Maximization of work extraction alone, a strategy typically employed in the information engine literature \cite{koski2014experimental, koski2015chip, camati2016experimental, paneru2018lossless, admon2018experimental, saha2021maximizing, mandal2013maxwell, barato2013autonomous, hoppenau2014energetics, cao2015thermodynamics}, is implemented by coarse graining. Our analysis shows that this strategy disregards non-negligible information processing costs and is, therefore, not generally thermodynamically rational. 
In a significant region of the problem's parameter space, viable engines that produce positive net work output, are possible only when the observer makes thermodynamically rational decisions. Even more intriguingly, the use of thermodynamic rationality reveals an interesting structure in parameter space, which cannot be discovered when the observable is coarse grained. 

\paragraph{Binary decisions and generalized \LS\ engines.} Solving a decision problem typically hinges upon knowledge of hidden variables, 
in the simplest case one random variable with binary outcome, $u\!=\!\pm1$. \LS's information engine embodies precisely that: A single-particle gas is prepared with a movable divider in the center of the container (at $x\!=\!0$) and connected to a heat bath at temperature $T$. An observer, who knows which side is empty, can extract work from the heat bath by isothermal expansion, up to, on average, $k T \ln(2)$ per engine cycle ($k$ denotes the Boltzmann constant). 
The observer's role is to store a transiently available observation in memory. A memory-dependent protocol on the piston is then executed mechanistically \cite{szilard-german}. The observer's freedom lies in the choice of the map from observation to memory.  
This mapping has to be implemented by a physical process, which requires energy. The minimum energy requirement depends on the map---storing more information comes at an inevitable cost \cite{CB, zurek1989algorithmic, sagawa2008minimal, parrondo2015thermodynamics}. 

For \LS's engine, the observable is the particle's $x$ position, the actionable variable, $u$, denotes which side is empty. By construction, the observer's choice does not pose much of a challenge: the optimal way to record to memory, $m$, is trivially given by the coarse graining $m = \sign(x)$, making $m$ identical to $u$. This decision problem is fully observable, since the available data always affords uncertainty free reconstruction of the actionable variable. \LS's engine converts work to information (when storing an observation in memory), and information to work (when exploiting the memory). Physical implementation of the map $m = \sign(x)$ requires, on average, at least $kT_M \ln(2)$, where $T_M$ denotes the temperature at which the memory making process runs. \LS's engine can therefore produce average net work output up to $k(T-T_M)\ln(2)$ per cycle. Run isothermally ($T=T_M$), the net output is zero, at best, and the engine acts purely as a converter between work and information. For $T_M < T$, \LS's engine is equivalent to a heat engine, the idealized process approaches Carnot efficiency \cite{stilldaimer2022}. This, however, assumes that the observer implements the decision strategy outlined above. There are, of course, other strategies that yield less work.

The decision problem becomes interesting when the hidden variable $u$ is not directly observable, and cannot be computed precisely from the accessible data. In the presence of intrinsic uncertainty, the challenge is to map data to memory in such a way that inference quality is high, while energetic costs remain low. Thermodynamically rational decision strategies can be probabilistic in this case, characterized by the conditional probabilities $p(m|x)$ \cite{CB}. When $m$ is given, there is an optimal, memory-dependent work extraction protocol, which can be applied mechanistically. It moves the divider into the direction that is most likely empty, leaving a residual volume fraction as large as the remaining inference error probability \cite{stilldaimer2022}. The observer's freedom in choosing a decision strategy thus lies in the choice of $p(m|x)$.

Decision problems with uncertainty can be linked to partially observable \LS\ engines in the following way. Importantly, this general correspondence makes our framework extendable to other types of uncertainty. For binary $u$, the intrinsic uncertainty is specified by the probability $p(u\!=\!1|x)$. A corresponding partially observable \LS\ engine can be constructed for any such problem with constant probability density $\rho(x) \!=\! \rho$ by re-ordering $x$, so that the $p(u\!=\!1|x)$ values are in ascending order, resulting in a monotonic function $d(x)$ \footnote{The construction for non-constant $\rho(x)$ is more involved, and discussed in \cite{co-sub}.}. A physical divider shaped as $d(x)$ is then used in the \LS\ engine, with any discontinuities connected vertically. The engine can be assumed, without loss of generality, to have a container of unit length in all directions. 

\paragraph{The problem considered.}

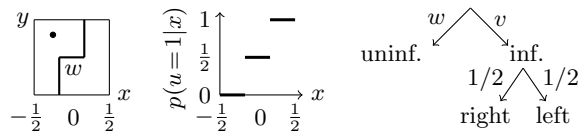
\begin{figure}
\vspace*{-0.3cm}
\hspace{-1.2cm}
\begin{minipage}[l]{0.3\linewidth}
    \begin{tikzpicture}
    \draw[black, thick] (1/3,0) -- (1/3,1/2);
    \draw[black, thick] (1/3,1/2) -- (2/3,1/2);
    \draw[black, thick] (2/3,1/2) -- (2/3,1);
    \draw[black] (0,0) -- (1,0);
    \draw[black] (0,0) -- (0,1);
    \draw[black] (0,1) -- (1,1);
    \draw[black] (1,0) -- (1,1);
    \node[] at (-0.2,1) {$\,\,\,\,$};
    \node[] at (0.6,1.6) {$\,$};
    \filldraw[black] (1/4,4/5) circle (1pt);
    \node[] at (1/1.9,1/3) {$w$};
    \node[] at (-0.1,-0.3) {$-\frac{1}{2}$};
    \node[] at (0.525,-0.3) {$0$};
    \node[] at (1,-0.3) {$\frac{1}{2}$};
    \node[] at (1.2,0) {$x$};
    \node[] at (-0.15,1) {$y$};
  \end{tikzpicture}
\end{minipage}
\hspace{-0.5cm}
\begin{minipage}[c]{0.3\linewidth}
    \begin{tikzpicture}
    \draw[->] (0,0) -- (1.1, 0) node[right] {$x$};
    \draw[->] (0, 0) -- (0, 1.1);
    \draw[black, line width=0.4mm] (0,0) -- (1/3,0);
    \draw[black, line width=0.4mm] (1/3,1/2) -- (2/3,1/2);
    \draw[black, line width=0.4mm] (2/3,1) -- (1,1);
    \node[] at (0.6,1.6) {$\,$};
    \node[] at (-0.2,0) {$0$};
    \node[] at (-0.2,1/2) {$\frac{1}{2}$};
    \node[] at (-0.2,1) {$1$};
    \node[] at (-0.1,-0.3) {$-\frac{1}{2}$};
    \node[] at (0.54,-0.3) {$0$};
    \node[] at (1,-0.3) {$\frac{1}{2}$};
    \node[rotate=90] at (-0.6, 0.5) {$p(u\!=\!1\vert x)$};
  \end{tikzpicture}
\end{minipage}
\hspace{0.2cm}
\begin{minipage}[l]{0.3\linewidth}
    \begin{tikzpicture}
    \draw[->, overlay] (0.0,0.6) -- (-0.5, 0.1);
    \draw[->, overlay] (0.0,0.6) -- (0.5, 0.1);
    \draw[->, overlay] (0.7,-0.2) -- (1, -0.64);
    \draw[->, overlay] (0.7,-0.2) -- (0.4, -0.64);
    \node[overlay] at (-1.05,0) {uninf.};
    \node[overlay] at (0.75,0) {inf.};
    \node[overlay] at (-0.45,0.45) {$w$};
    \node[overlay] at (0.4,0.43) {$v$};
    \node[overlay] at (0.2,-0.35) {$1/2$};
    \node[overlay] at (1.2,-0.35) {$1/2$};
    \node[overlay] at (0.2,-0.85) {right};
    \node[overlay] at (1.1,-0.82) {left};
  \end{tikzpicture}
\end{minipage}
\caption{The decision problem. Left: as realized by a partially observable \LS\ engine; center: specified by the conditional probability $p(u=1 \vert x)$. Right: a fraction $v$ of observable outcomes are informative, of those, half correspond to the left side being empty; $w=1-v$ are uninformative.
}
\label{Fig:WorkMedium_Memory}
\end{figure}

Here, we consider a most basic decision problem: a fraction $v \in (0, 1)$ of the observable outcomes $x$ tell us with certainty which side is empty, i.e. $p(u|x)$ is either 0 or 1 for these $x$. The rest of the observations tell us nothing at all about $u$, i.e., those $x$ have $p(u|x)=1/2$.
The corresponding partially observable \LS\ engine has a divider with a horizontal part of width $w=1-v$ in the middle (at $y=0$), and two vertical parts, located at $x=\pm w/2$ (left panel of Fig. \ref{Fig:WorkMedium_Memory}). 
The problem is characterized by
\begin{equation}
p(u\!=\!1|x) = \left\{\begin{array}{ll} 0 & {\rm if}\, x\in {\cal X}_L = [-1/2, -w/2] \\ 1/2 & {\rm if} \, x\in {\cal X}_M = (-w/2, w/2) \\ 1 & {\rm if}\, x\in {\cal X}_R = [w/2, 1/2] \end{array}\right.,
\end{equation}
(Fig. \ref{Fig:WorkMedium_Memory}, center). 
It contains $v$ bits of usable information:
$I[X,U] = \left\langle \left\langle \ln\left[ p(u\vert x) / p(u) \right]\right\rangle_{p(u|x)}\right\rangle_{\!\rho(x)} \!=\! v \ln(2)$.

As in \LS's original engine, the divider can be used as a piston to extract work by increasing the accessible volume. With access to the particle's $x$-position alone, the accessible volume can, at best, be doubled in a fraction $v$ of all engine cycles.

\paragraph{Naive coarse graining.}
The maximum average amount of work, $k T v \ln(2)$,
can be extracted if the observer coarse grains $x$ into the three different regions, ${\cal X}_L$, ${\cal X}_M$, and ${\cal X}_R$. Coarse graining is a data representation strategy that implements a hard partition of the observable space by using a deterministic map, in this case mapping to three distinct memory states,
\begin{equation}\label{Eq-mofx-3}
    \!\!p(m|x) = \delta_{m \widehat{m}(x)},~{\rm with} \;\; \widehat{m}(x) = 
\left\{\begin{array}{rcl}  
-1 & \text{if}& x \in \mathcal{X}_L \\ 
0 & {\text{if}}& x \in \mathcal{X}_M ,\\
1 & {\text{if}}& x \in \mathcal{X}_R
\end{array} 
 \right. 
\end{equation}
where $\delta_{ab} = 1$ if $a=b$, and $\delta_{ab} = 0$ otherwise. 
This classification embodies two binary decisions: first, decide if the data are informative or not, then, if informative, decide if they belong to ${\cal X}_L$ or to ${\cal X}_R$. The situation is sketched in the diagram on the right of Fig. \ref{Fig:WorkMedium_Memory}. As a consequence of the deterministic nature of this partitioning, the conditional entropy is zero, $\widehat{H}[M|X]=0$, and the total amount of information kept in memory is $\widehat{I}[M,X] = \widehat{H}[M]$. To evaluate the entropy, we introduce 
$h(x) :=\!- \!\left(1\!-\! x \right) \ln\left(1\!-\!x\right) - x \ln(x) \!>\! 0$ for $0 \!<\! x\! <\! 1$, $h(0)=h(1)=0$. The choice of the base of the logarithm determines the units of information; for base 2 we write $h_2(x) = h(x)/\ln(2)$, and we have $\widehat{H}_2[M] = \widehat{H}[M]/\ln(2) = h_2(v) + v$. There are $h_2(v)$ bits of information about whether or not a data sample is informative. But this is useless information for energy harvesting purposes. Only $v$ bits of the total memory reveal {\em which} side of the container is empty, and are therefore usable. Coarse graining into three states is wasteful, because it keeps $h_2(v)$ unusable bits. 

Strategies such as this one, which maximize gain regardless of cost, are usually used in the information engine literature. These strategies are not always thermodynamically rational. To see how bad this type of strategy can be, note that, in this example, using Eq. (\ref{Eq-mofx-3}) leads to an upper bound on net average work output of $\widehat{W}_{\rm out}^{\rm net} = k (T - T_M) v \ln(2) - k T_M h(v)$. This bound is {\it negative} when the relative temperature difference, $\Delta \tau := (T-T_M)/T_M$, is below ${h_2(v)}/{v}$. Typical experimental realizations have $T=T_M$. That would imply an average {\it net work loss} of, at best,
$k T h(v)$ per cycle, if no other dissipation was encountered in the implementation. The observer would be better off doing nothing at all. This shows that data processing costs cannot be ignored, and at least a lower bound should be compared to the derived gain in experimental studies.
 
Maximizing energetic gain alone, regardless of processing costs, is thermodynamically rational only for $T >> T_M$, because in this regime energetic costs associated with keeping more information become negligible, as every usable bit allows for comparably enormous gain.
But the engine's behavior is then dominated by the large temperature gradient. In the more interesting regime of small temperature ratios, $\tau := T/T_M$, the quality of the observer's information processing governs the engine's behavior. In this regime rational decisions make a considerable difference.

\paragraph{Thermodynamic rationality.} 
For any given observer strategy, $p(m|x)$, the upper limit on average net engine work output is the maximally possible work gain, $kTI[M,U]$, minus the smallest possible amount of work needed to run the memory, $kT_M I[M,X]$ \cite{CB}:
\begin{equation}
W_{\rm out}^{\rm net} = k (T I[M,U] - T_MI[M,X])~.   
\end{equation}
Thermodynamically rational agents use that strategy $p^*(m|x)$, which maximizes $W_{\rm out}^{\rm net}$.
Mathematically, such strategies are solutions to the following optimization problem: ${\rm arg}\!\max_{p(m|x)} W_{\rm out}^{\rm net}[p(m|x)]$; subject to $\sum_m p(m|x) = 1$, $\forall x$ \cite{CB}. 
These solutions are computable numerically with the ``Information Bottleneck" algorithm \cite{IB}. 

When the observer remembers something, we have \mbox{$I[M,X]\!>\!0$}. The ratio of maximally extractable work to minimal cost then cannot exceed $kTI[M,U] / kT_MI[M,X] \leq \tau$, since $I[M,U] \leq I[M,X]$. The temperature ratio $\tau$ determines the {\em thermodynamic value} of usable information. 
We expect the observer to remember varying degrees of detail, depending on $\tau$, because keeping more information requires more work, an investment that, in general, makes sense only if the work that can be gained from the usable part of the memorized information, $I_{\rm u} := I[M,U]$, exceeds the invested amount. The rest of the total information captured, $I_{\rm m} := I[M,X]$,  is useless: $I_{\rm ul} := I_{\rm m} - I_{\rm u}$. 

The optimization problem can equivalently be understood as finding the data representation that contains the least amount of useless information, subject to capturing as much useful information as possible: i.e. finding \mbox{${\rm arg}\!\min_{p(m|x)} (I_{\rm ul}[p(m|x)] - \Delta \tau I_{\rm u}[p(m|x)])$}; s.t. $\sum_m p(m|x) = 1$. The relative temperature difference, $\Delta \tau = \tau-1$, controls the trade-off.  

We saw that, in the range $1< \tau < 1+ h_2(v) / v$, strategy Eq. (\ref{Eq-mofx-3}) is worse than doing nothing. But we also know that for $ \tau > 1$, the engine could in principle produce positive average net work output, as long as the observer can find a memory that captures enough usable information without being too wasteful. What do rational decision strategies look like in this range? Do they rely on two memory states, or on three? 

\begin{figure}[ht]
\centering
    \includegraphics[width=\linewidth]{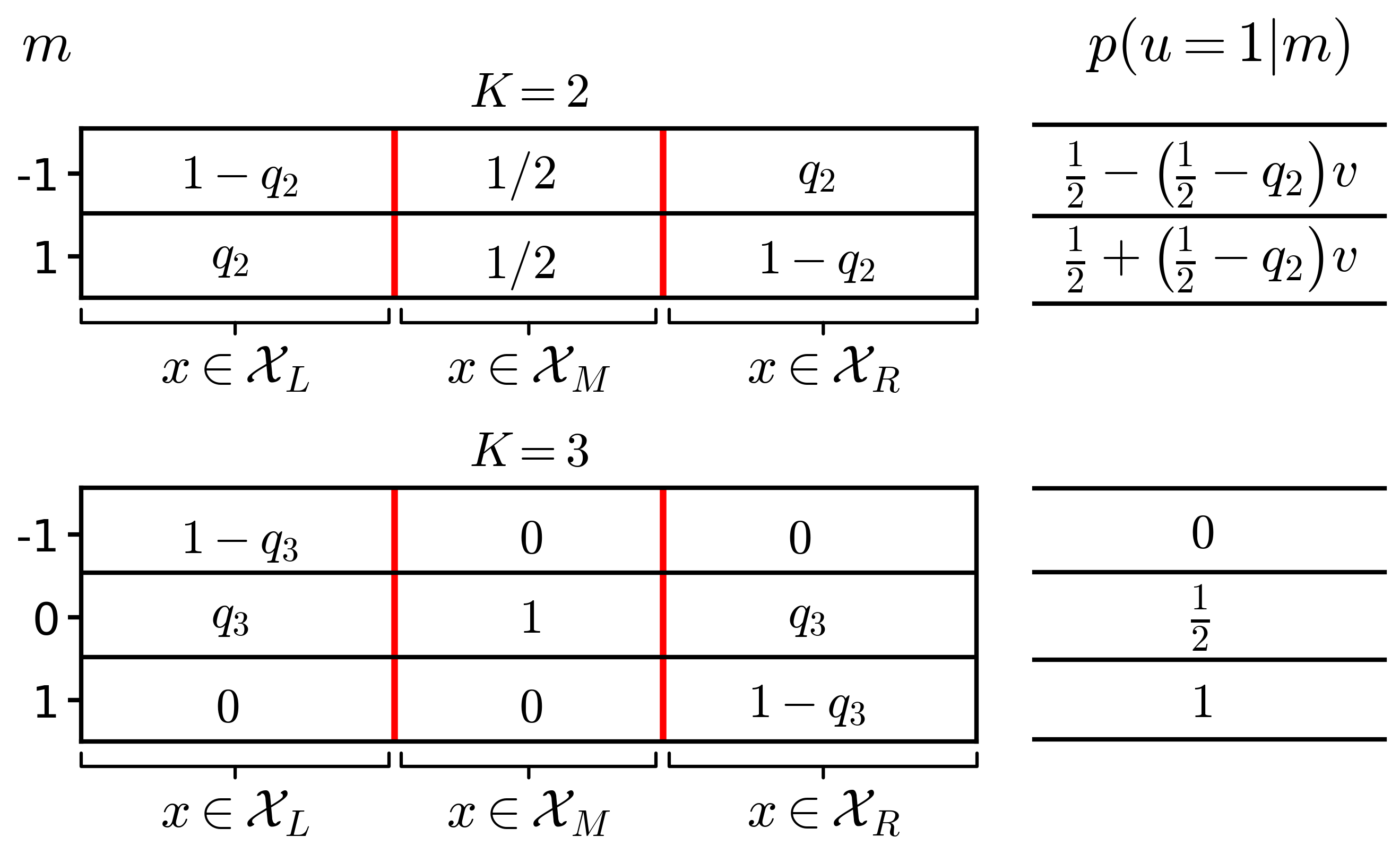}
    \label{Fig:Approximations}
    \vspace{-2em}
\caption{Thermodynamically rational decision strategies are soft partitions, characterized by $p(m|x)$ (displayed in the cells, left), parameterized by the number of memory states, $K(v, \tau)$, and the probabilities $q_2(v, \tau)$ and $q_3(v, \tau)$. Red lines delineate the region with uninformative outcomes, shown is $v=0.7$. Right: Each memory state $m$ carries an inference of $u$.
}
\label{Fig:parametric-solutions}
\end{figure}

To answer these questions we enlisted the help of the algorithm \cite{co-sub}, but visual inspection of the numerical solutions revealed a parametrization of thermodynamically rational observer strategies (shown in \mbox{Fig. \ref{Fig:parametric-solutions}) \footnote{We verified numerically by brute force that this parametrization yields optimal engine performance.}.}

This parametrization enables analytic characterization of thermodynamically rational agent behavior. The critical $\tau$ values at which transitions occur between one, two, and three distinct memory states can be calculated. To that end, we write the average net engine work output, $W_{\rm out}^{\rm net}$, as a function of $\tau$, $v$, and $q_K$. Optimal $q_K(v,\tau)$ are then determined from the condition \mbox{$d W_{\rm out}^{\rm net}/d q_K = 0$}, yielding \mbox{$q_3 = (1-v)/v(2^{\Delta \tau}-1)$}, and a transcendental equation for $q_2$ \cite{co-sub}, which has an analytic solution for $\tau = 2$: \mbox{$q_2 = \frac{1}{2} - \frac{1}{2v}\sqrt{2v-1}$} (for $v \geq 1/2$). Critical values for $\tau$ are obtained by analytical comparison of $W_{\rm out}^{\rm net}$ for different numbers of underlying memory states, $K \in \{1,2,3\}$.

We find that two-state strategies emerge only if the majority of observation outcomes are informative ($v > 1/2$). In that case, the transition from one to two memory states occurs at the critical value
\begin{equation}
\tau_{1\rightarrow 2}(v) = \frac{1}{v}~, ~~~ v \in \left(1/2, 1\right),
\label{tau12}
\end{equation}
and the transition from two to three memory states occurs
universally ($\forall\, v > 1/2$) at 
\begin{equation}
\tau_{2\rightarrow 3} = 2~. \label{tau23}
\end{equation} 

Thermodynamically rational strategies in the range $\tau_{1\rightarrow 2}(v) < \tau < \tau_{2\rightarrow 3}$ use two memory states as displayed in the upper panel of Fig. \ref{Fig:parametric-solutions}: 
Any observation that is fully informative is assigned with high probability, $1-q_2$ ($q_2<1/2$), to one of those memory states. The value $m$ is just a label, and we are free to choose the labeling convention such that $x\in {\cal X}_L$ get assigned to $m=-1$, and $x\in {\cal X}_R$ to $m=1$. Uninformative observations in the middle region are assigned to either state at random, requiring no work. The energetic costs are thereby reduced without negatively affecting inference quality \cite{co-sub}. 

Using this two-state memory allows the agent to extract work up to $kT\ln(2) \left[1 - h_2\left(\frac{1}{2} + \big(\frac{1}{2}-q_2\big)v\right)\right]$, at a minimum energetic cost of $kT_M \ln(2) v \big[1-h_2(q_2)\big]$.
Reducing costs becomes more important with smaller $\tau$, as usable information has less thermodynamic value. Cost reduction is implemented by softening the partition: $q_2(v,\tau)$ monotonically increases with decreasing $\tau$. 

At $\tau_{2\rightarrow 3}=2$ (and $v > 1/2$), $W_{\rm out}^{\rm opt} = W_{\rm out}^{\rm net}|_{p^*(m|x)}$ is readily calculated for two- and three-state memories. It is $\frac{1}{2} kT \ln(2) [ 1 - h_2(v)]$ in either case, explaining the transition. Comparison to the naive coarse graining strategy, which gives $\widehat{W}_{\rm out}^{\rm net} = \frac{1}{2} k T \ln(2) [v - h_2(v)]$ for $\tau=2$, reveals that thermodynamic rationality results in an advantage of $W_{\rm out}^{\rm opt} - \widehat{W}_{\rm out}^{\rm net} = \frac{1}{2} kT \ln(2) [ 1 - v]$.

For $\tau > 2$, three-state memories (lower panel of Fig. \ref{Fig:parametric-solutions}) yield better engine performance than two-state memories. The data representation strategy changes dramatically: with two states, uninformative observations are assigned at random, while with three states they are mapped with probability one to the third state, i.e. resources are dedicated to being certain about knowing nothing. Observations in the regions ${\cal X}_L$ and ${\cal X}_R$ are mapped to the respective memory state with probability $1-q_3$, and with probability $q_3$ to the new $m=0$ state. This is a clever way to save coding costs by softening the partition, while not compromising the inference accuracy for memory states $m=\pm 1$, which is maximal: $p(u\!=\!m|m\!=\!\pm1) \!=\! 1$. The observer can thus extract average work up to $kT \ln(2)$ whenever $m=\pm 1$ is recorded. The probability of this occurrence is \mbox{$p_{\pm} := \sum_{m =\pm 1} p(m) = v(1-q_3) = (v 2^{\Delta \tau}-1)/(2^{\Delta \tau}-1)$}. 

As $\tau$ increases, $p_{\pm}$ increases, whereby the observer can profit more often from their perfect inference. The maximum extractable work is $kT\ln(2) p_{\pm}$, and the minimum coding cost is \mbox{$kT_M \ln(2) \big[ p_{\pm} + h_2(p_{\pm}) - v h_2(p_{\pm}/v)\big]$}. In the limit of large $\tau$, when $p_{\pm} \rightarrow v$, the strategy approaches Eq. (\ref{Eq-mofx-3}).
The advantage of thermodynamically rational three-state strategies over the naive coarse graining of Eq. (\ref{Eq-mofx-3}), in terms of the difference in the respective $W_{\rm out}^{\rm net}$, is linear in the fraction of uninformative samples, and decreases monotonically with $\tau$: $W_{\rm out}^{\rm opt} - \widehat{W}_{\rm out}^{\rm net} = kT_M (1-v)\ln[2^{\Delta\tau}/(2^{\Delta\tau} - 1)]$. 

If uninformative outcomes dominate ($v \leq 1/2)$ then thermodynamically rational observers do not profit from utilizing two memory states at any value of $\tau$. Instead, they either remember nothing and do nothing (one-state memory), or they use a three-state memory. This is the case when the relative temperature difference exceeds the self-information \cite{hsiao1957processing, cover1999elements} of an informative outcome, measured in bits: $\Delta \tau > \log_2\left(1/v \right)$. That is, the transition from one to three memory states occurs at
\begin{equation}
\tau_{1\rightarrow 3}(v) = 1 + \log_2\left(\frac{1}{v}\right)~, ~~ v \in \left(0, 1/2\right]~. \label{tau13}
\end{equation}

\paragraph{Summary of results.}
These analytical results show how the thermodynamic value of information dictates the appropriate detail with which information ought to be stored---in other words, it dictates the appropriate complexity of the summary an observer makes of available data.
Recall that, even when data are abundant, and statistical overfitting \cite{dietterich1995overfitting} is not an issue, an observer's model can still be overly complicated. For partially observable information engines model complexity is physical: the model should not retain information that the observer cannot gainfully utilize.

\begin{figure}[ht]
    \centering
    \includegraphics[width=\linewidth]{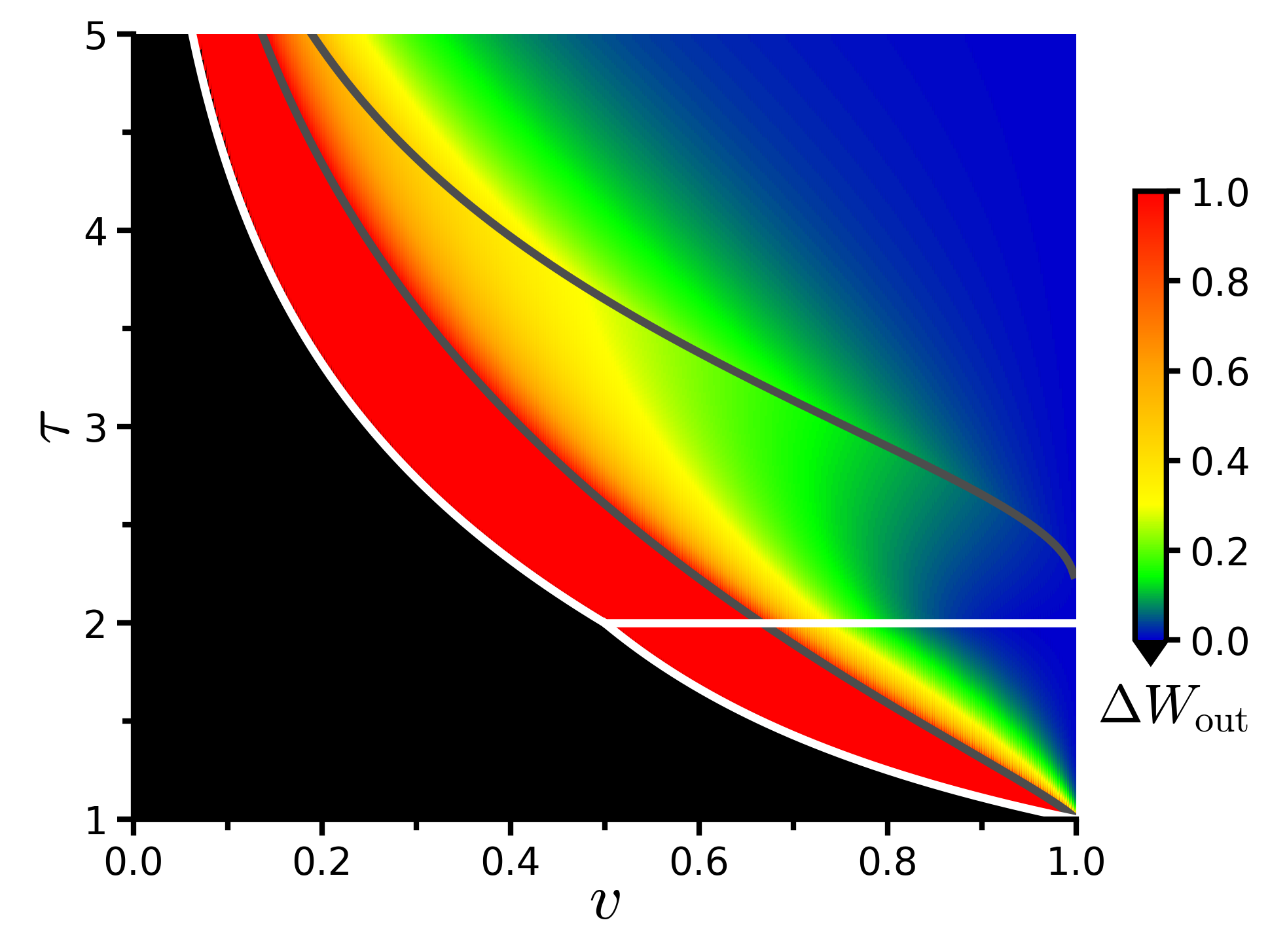}
    \vspace{-2em}
\caption{Phase diagrams comparing engines run by thermodynamically rational agents (white lines, Eqs. (\ref{tau12}) - (\ref{tau13})) to engines run by observers restricted to coarse graining (grey lines, Eqs. (\ref{tau12det}) - (\ref{tau23det})), overlaid with quantitative performance difference, $\Delta W_{\rm out}$, in color (see text). In the red area, thermodynamic rationality results in positive net engine work output, while coarse graining does not. Displayed is $v < 1$ ($v=1$ is \LS's engine without uncertainty).}
\label{Fig:Heatmap}
\end{figure}

Results are summarized in the phase diagram in $v$--$\tau$ parameter space, shown in Fig. \ref{Fig:Heatmap}. White lines, Eqs. (\ref{tau12})-(\ref{tau13}), delineate regions in parameter space in which thermodynamically rational agents use one, two, or three memory states, respectively. In the black area below the white lines the engine cannot produce positive average net work output, and the best strategy is to do nothing (corresponding to one memory state). Viable engines are found only in the colored areas. The colors compare to the performance of the best coarse graining strategy.

\paragraph{Observers that can only coarse grain.}
Coarse graining observables is a standard method to think about the physics of stochastic and complex systems \cite{gibbs1902elementary, ehrenfest1907begriffliche}. It is interesting to note that observers that coarse grain are not generally implementing thermodynamic rationality.
The color in Fig. \ref{Fig:Heatmap} shows the performance difference between what a thermodynamically rational observer can achieve in terms of average net engine work output, $W_{\rm out}^{\rm opt}$, compared to what an observer that uses minimally dissipative coarse graining can achieve, $W_{\rm out}^{\rm det}$. Displayed is the difference in units of the optimal output: $\Delta W_{\rm out} := (W_{\rm out}^{\rm opt} - W_{\rm out}^{\rm det}) / W_{\rm out}^{\rm opt}$ \footnote{Obtained from numerical calculations.}. 

In the red area, thermodynamically rational agents can operate the engine, producing positive net work output, while observers limited to coarse graining cannot. 
This area dominates in the regime of main interest, in which temperature differences are not large, and uncertainty is not negligible, but also not overwhelming. For $0.5< v \leq 0.9$ and $\tau \leq 2$, the red area covers over 60\% of the total feasible area. The performance difference becomes negligible only in the upper right corner (blue color).

The following strategies are used by observers that maximize $W_{\rm out}^{\rm net}$ under the coarse graining constraint: Eq. (\ref{Eq-mofx-3}) at large $\tau$; coarse graining into two regions with boundary at either $-w/2$ or $w/2$ at intermediate $\tau$. The latter strategy lumps uninformative observable outcomes together with (either) one of the informative outcome regions \cite{co-sub} (those two solutions are degenerate). Critical $\tau$ values are readily calculated. Transitions occur at
\begin{eqnarray}
 \tau^{\rm cg}_{1 \rightarrow 2}(v) \!&=&\! \frac{h\!\left({v \over 2}\right)}{\ln(2) \!-\! \left(1\! -\! \frac{v}{2}\right) h\!\left(\frac{1}{2-v}\right)}~, \label{tau12det}  \\
 \tau^{\rm cg}_{2 \rightarrow 3}(v) \!\!&=&\!\! 1 \!+\! \frac{2v\ln(2) - h(v)}{(1\!-\!v)\ln(4(2\!+\!v^2\!-\!3v))\!-\!\ln(2\!-\!v)}~. \label{tau23det}     
\end{eqnarray}
These functions are displayed as dark grey lines in \mbox{Fig. \ref{Fig:Heatmap}} ($\tau^{\rm cg}_{1 \rightarrow 2}(v) < \tau^{\rm cg}_{2 \rightarrow 3}(v)$), showing that the phase diagram is markedly different when the observer coarse grains.

Importantly, the qualitative change we observed for thermodynamically rational strategies, delineating whether the majority of possible observation outcomes is informative ($v > 1/2$) or not ($v \leq 1/2$), disappears. Instead, from the observer's point of view, it now appears that there is a region in which two states are preferable for any $v$. The universality of the second phase transition (Eq. (\ref{tau23})) also disappears. Thus, when the observer is restricted to coarse graining, the interesting structure in the problem cannot be discovered.

\paragraph{Conclusion.}
We showed that the physics of rational decision making under uncertainty can be completely understood and characterized analytically for  a simple, yet non-trivial, Maxwell's demon with access to data of which only a fraction is informative. 

Our analysis showcases the thermodynamic foundations of decision making under uncertainty for a fundamental decision problem: the observer's actions are based on a {\it binary} underlying variable that is inferred from incomplete information. Coarse-graining is no longer the observer's go-to method. Instead, more subtle strategies are rational. 
Our results provide an important design rule for thermodynamically efficient computing machinery. Beyond that, there may be wider implications for physical modeling in other domains where coarse graining is ubiquitous. Our transparent example also exposes the fact that energetic costs for information processing should not be ignored in the analysis of information engines, as they can be substantial.

How much trust can be put in models and methods heavily dependent on {\it ad hoc} choices, as are found in domains such as machine learning, decision theory and complex systems modeling? Those choices rely on the designer's intuition, and we showed here that even for this simplest of cases intuition might not be clear cut when it comes to rational decision making under uncertainty. 

\paragraph{Acknowledgments.}
\noindent We thank Rob Shaw for helpful comments on the manuscript. We are most grateful for funding from the Foundational Questions Institute, Grant Nos. FQXi-RFP-1820 (FQXi together with the Fetzer Franklin Fund) and FQXi-IAF19-02-S1. This publication was made possible through the support of the ID\# 62312 grant from the John Templeton Foundation, as part of the \href{https://www.templeton.org/grant/the-quantum-information-structure-of-spacetime-qiss-second-phase}{‘The Quantum Information Structure of Spacetime’ Project (QISS)}. The opinions expressed in this publication are those of the author(s) and do not necessarily reflect the views of the John Templeton Foundation.

\bibliographystyle{unsrt}
\bibliography{DecisionMakingUnderUncertainty}
\end{document}